\documentclass[final,5p,times,twocolumn]{elsarticle}

\usepackage{amssymb}

\journal{Journal of Quantitative Spectroscopy and Radiative Transfer}

\def\Re{{\rm Re\mit}}
\def\Im{{\rm Im\mit}}

\begin{document}

\begin{frontmatter}



\title{Enhanced energy transfer via graphene--coated wire surface plasmons}

\author[label1]{Julieta Olivo} 
\author[label2,label3]{Mauro Cuevas \corref{cor1}}
\ead{cuevas@df.uba.ar}
\address[label1]{Universidad Nacional de R\'io Cuarto, Dto. F\'isica, Facultad de Ciencias Exactas Fisicoqu\'imicas y Naturales, R\'io Cuarto (5800), C\'ordoba, Argentina}
\address[label2]{Consejo Nacional de Investigaciones Cient\'ificas y T\'ecnicas (CONICET) and Facultad de Ingenier\'ia y Tecnolog\'ia Inform\'atica, Universidad de Belgrano,
Villanueva 1324, C1426BMJ, Buenos Aires, Argentina}
\address[label3]{Grupo de Electromagnetismo Aplicado, Departamento de F\'isica, FCEN, Universidad de Buenos Aires and IFIBA, Ciudad Universitaria, Pabell\'on I, C1428EHA, Buenos Aires, Argentina}

\cortext[cor1]{corresponding author}
\author{}

\address{}

\begin{abstract}
This work analyzes  the electromagnetic energy transfer rate between donor and acceptor quantum emitters close to a graphene--coated wire. We discuss the  modification of the energy transfer rate when the emitters  are interfaced via surface plasmon (SP) environments. All of the notable effects on the spatial dependence of the energy transfer are highlighted and discussed in terms of SP propagation characteristics. Our results show that a  dramatically enhancement of the energy transfer occur when the graphene wire SPs are excited. Moreover, different dipole moment orientations influence differently this enhancement.   As a consequence of the
quasi--one--dimensional graphene wire SPs,  we found  that the normalized energy transfer rate reaches a maximum value at a donor--acceptor distance which is twice the value corresponding to its two--dimensional counterpart consisting of a single graphene sheet or a flat graphene waveguide. In particular, we provide a simplified model that reproduces the main features of the numerical results.
\end{abstract}


\end{frontmatter}



\section{Introduction}

The potential of SPs to confine the electromagnetic field to regions well below the diffraction limit has been widely exploited  to manipulate the light--matter interaction at subwavelength scales enabling a rich variety of applications \cite{maier_nature,review,dereux}. In particular, the coupling of optical emitters, such as atoms, molecules or quantum dots, to SP modes provides an  enhancement of the emission rate several orders of magnitude relative to the case in which the same emitter is localized in an unbounded medium. This property has been used to bring the size of laser sources of light to the nanoscale, to reduce the lasing threshold \cite{laser,laser1} as well as to improve the efficiency of single photon sources \cite{SPS}.   
In addition, SP mode environments can provide an enhancement together a significant directional control of the energy transfer between quantum emitters over distances larger than the F\"orster energy transfer range \cite{forster,barnes,vasili1,ETMM,Jones}.    

Significant progress made in nanoscale fabrication techniques and an extensive wealth of theoretical analysis have not only allowed an advanced light manipulation via SPs at optical frequencies (where optical emitters are placed on metallic SP environments),  but have also led to a rapidly developing field in light control by plasmonic excitation in the terahertz (THz) frequency range \cite{THz}. 
Possibilities have been widened with the advent of other plasmonic materials with lower losses and greater confinement of the electromagnetic field, such as metal--alloys, heavily doped wide--band semiconductors, and graphene \cite{shalaev}. Due to their unique  property to guide SPs by one--atom--thin layer from microwaves to the mid--infrared regimes \cite{Xia,nikitin10,nikitin100}, graphene has attracted significant  interest from the nanophotonics research community.    
The electronic linear band structure of graphene makes a SP mass depending on the Fermi--level position and consequently electrically tunable and long--lived SPs are supported by graphene. 

One of the major challenges for most of applications is to build more smaller plasmonic constituent elements to control light at subwavelength scales,  in particular from microwaves to the mid--infrared regimes which comprises a crossover between electronic and  optics. In this way, graphene--plasmonic has found a great variety of   
sophisticated applications, such as photonic devices capable to achieve invisibility \cite{Naserpour,Tuz},  electrically tunable THz antennas \cite{jornet,filter,gomez_diaz,cuevas6} and efficient THz waveguides \cite{depineguias}. 
The propagation characteristics of graphene plasmonic waveguides 
have been investigated in different geometrical  structures, such as planar waveguides \cite{marocico,shadrivov,cuevas0} ribbons \cite{nikitinguias,christensenguias}, grooves \cite{liuguias} and graphene--cylinders with circular cross--section \cite{enguetaguias}. 
The ability of plasmonic graphene wires to guide light confined well below the diffraction limit has the key advantage for tailoring light--matter interactions \cite{vasili2,martinmoreno}, opening routes for novel electrically controllable devises capable to improve the single emitter emission as well as the energy transfer rate between optical emitters \cite{biehs, shahbazyan2,arruda2,cuevas3,ETesfera,cuevas3bis}.  

Methods for controlling the relative positions of single emitters and SP wires with high experimental accuracy have been reported.  In \cite{gruber}, it was described the use of  electron beam lithography to create a polymer template that enables the controlled deposition of a small number of quantum dots in areas that measure about 50nm across.  Alternatively, in \cite{roop}  the authors have reported the use of a microfluidic device for positioning at desired locations  and moving quantum dots  around a single metallic nanowire with a 12nm spatial accuracy. 

This paper deals with the study about the energy transfer rate between single emitters, a donor and acceptor,  located near a graphene--coated wire. In a recent work \cite{cuevas4}, we have revealed that the interplay between an optical emitter and SP excitations on a graphene layer strongly influences the energy transfer process between two single optical emitters placed close to a graphene--coated cylinder. However, in that work we have focused on a two--dimensional model in which each of the optical emitters is considered as an infinitely thin wire extending  along the $z$ axis of the cylinder and with its dipole moment in the computational $x-y$ plane. As a consequence, only localized surface plasmons, \textit{i.e.}, SPs whose  wave vector component along the wire axis is equal to zero, are excited, leading to an optical energy  transfer from donor to acceptor within the $x-y$ plane.     

Our motivation for this work is to extend the study realized in \cite{cuevas4} to the case of quantum dot optical emitters. Unlike the two--dimensional approach, SPs propagating along the waveguiding structure, \textit{i.e.}, SPs  with a non null wave vector component along the wire axis, are excited and consequently  the energy can be transferred  between two distant optical  emitters placed near the graphene--wire at a distance $\Delta z$ from each other. In this sense, the present work can be considered as an extension to the cylindrical geometry case of our previous study about the effect of SPs on the resonant energy transfer on  a planar graphene waveguide \cite{olivo}. 

This paper is organized as follows. In section 2, 
we sketch the boundary--value problem for the diffraction problem  of a dipole emitter  located in close proximity to a graphene--coated wire. Because of the translational symmetry of the system along the direction of the wire axis, the transverse components of the electromagnetic fields are obtained from their longitudinal components for which we derive integral expressions for the scattered fields. We then include a second dipole emitter and deal with the problem of the coupled system. 
By using contour integration in the complex plane, we have applied a method developed in Ref. \cite{olivo}   to perform the field integration. This method which is  based on the residues theorem enables a fast calculation of the contribution of each one of the wire  modes to the energy transfer rate.   In section 3 
we present numerical results obtained under different dipole moment configurations. Concluding remarks are provided in Section 4. 
The Gaussian system of units is used and an $\mbox{exp}(-i\, \omega\, t)$ time--dependence is implicit throughout the paper, where $\omega$ is the angular frequency, $t$ is the time coordinate, and $i=\sqrt{-1}$. The symbols Re and Im are used for denoting the real and imaginary parts of a complex quantity, respectively.

\section{Theory} \label{teoria} 

We consider the energy transfer rate between a donor D and an acceptor A electric dipoles placed close to a  graphene--coated wire, as illustrated in Figure 1. The energy transfer  in the presence of the graphene--coated wire normalized to that in an unbounded medium 1 (without graphene waveguide), is given by \cite{novotny}
\begin{eqnarray}\label{ETn}
F_{ET}=\frac{|\hat{n}_A\cdot \textbf{E}_D(\textbf{r}_A)|^2}{|\hat{n}_A\cdot \textbf{E}_0(\textbf{r}_A)|^2}, 
\end{eqnarray}
where  the electric field $\textbf{E}_D(\textbf{r})$   is the electric field in the presence of a graphene--coated wire (see \ref{ap2}), $\textbf{E}_0(\textbf{r})$ is due to an electric dipole in an unbounded medium (see \ref{ap1}),  $\hat{n}_A$ is a unit vector along the induced polarization of the acceptor (whose direction is assumed to be fixed) and $\textbf{r}_A$ is the position of the acceptor.

\begin{figure}
\centering
\resizebox{0.5\textwidth}{!}
{\includegraphics{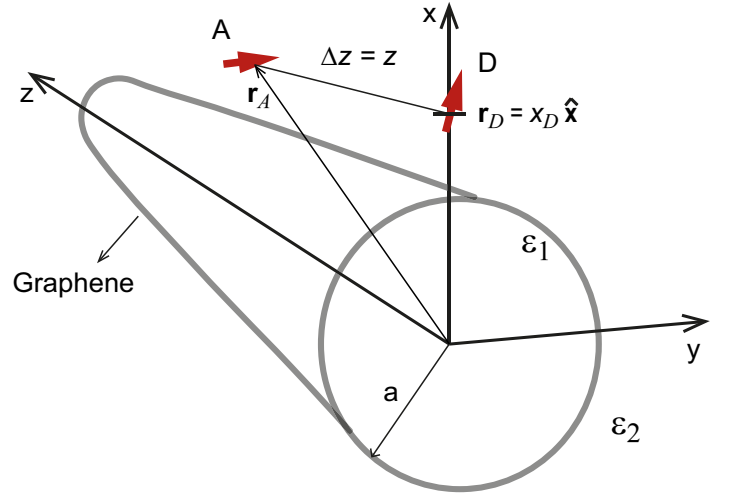}}
\caption{\label{fig:epsart} Schematic illustration of the graphene--coated wire. Both donor and acceptor emitters are placed outside the graphene--wire at $\textbf{r}_D=\rho_D \hat{x}$ and $\textbf{r}_A=\rho_A \hat{\rho}+\phi_A \hat{\phi}+z\hat{z}$, respectively. The graphene--wire ($\varepsilon_1=2.13$, $\mu_1=1$ and surface conductivity $\sigma$) is embedded in a transparent medium with constitutive parameters $\varepsilon_2$, $\mu_2=1$. 
}\label{grafeno}
\end{figure}

We consider the case in which the donor is placed outside the graphene--coated wire ($\rho_D>a$). 
Without loss of generality we choose $\phi_D=0, \, z_D=0$, \textit{i.e.} the donor position $\textbf{r}_D=\rho_D \hat{x}$ as shown in Figure 1.  
By using Eqs. (\ref{campos_transversales}), (\ref{E2z}) and (\ref{H2z}) we obtain the transverse components of the scattered electric field for $\rho>a$,
\begin{eqnarray}\label{camporho}
E^{(2)}_{\rho}(\textbf{r})=
\sum_{m=-\infty}^{+\infty}\int_{-\infty}^{+\infty} dk_z e^{i m \phi} e^{i k_z z} \nonumber\\
\times \left[
-\frac{k_0 m}{\gamma^{(2)}} \frac{H_m(\gamma^{(2)}\rho)}{\gamma^{(2)}\rho} a_m^{(2)} + i \frac{k_z}{\gamma^{(2)}} H'_m(\gamma^{(2)}\rho) b_m^{(2)}\right],    
\end{eqnarray}
and 
\begin{eqnarray}\label{campophi}
{E}^{(2)}_{\phi}(\textbf{r})=
-\sum_{m=-\infty}^{+\infty}\int_{-\infty}^{+\infty} dk_z e^{i m \phi} e^{i k_z z} \nonumber \\
\times \left[
i\frac{k_0}{\gamma^{(2)}} H'_m(\gamma^{(2)}\rho) a_m^{(2)} +  \frac{k_z m}{\gamma^{(2)}} \frac{H_m(\gamma^{(2)}\rho)}{\gamma^{(2)} \rho} b_m^{(2)}\right],
\end{eqnarray}
where the complex amplitudes $a_m^{(2)}$ and $b_m^{(2)}$ are given by expressions (\ref{am}) and (\ref{bm}), respectively. 
The integration path in Eqs. (\ref{camporho}) and (\ref{campophi}) is set along the real $k_z$ axis and  
poles, i.e., zeroes of the denominator  in $a_m^{(2)}$ and $b_m^{(2)}$ amplitudes, close to that axis give  rise to poor numerical convergence. To avoid this difficulty we have applied a method developed in Ref. \cite{olivo} to perform such integration which requires the application of the residues theorem to  extract each pole contribution of the integrals (\ref{camporho}) and (\ref{campophi}). 
%
\begin{figure}
\centering
\resizebox{0.5\textwidth}{!}
{\includegraphics{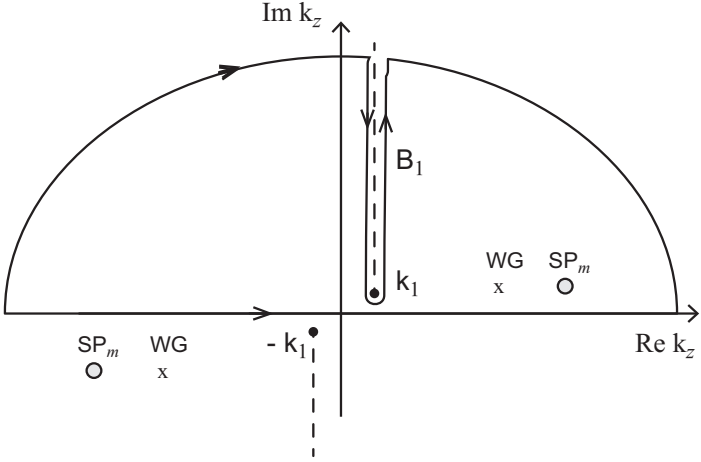}}
\caption{\label{fig:epsart} Singularities and path of integration in the complex plane
$k_z=\Re\,k_z+i\,\Im\,k_z$ for the electromagnetic fields. Pole singularities representing the propagation constant of the eigenmodes supported by the graphene--coated wire, like waveguide (WG) or surface plasmon (SP) modes, are illustrated with crosses and circles, respectively. The vertical lines drawn from the branch point $k_1=k_0 \sqrt{\varepsilon_1}$ to $+i\infty$ ($-k_1$ to $-i\infty$) are the branch cut lines.  Integration path with vertical branch cuts and poles captured. 
}\label{figura2}
\end{figure}
%
In order to perform the  calculation of  these pole contributions  to the field,  
we deform the integration path in (\ref{camporho}) and (\ref{campophi}) into a semicircle of large radius ($|k_z|\rightarrow \infty$) avoiding the branch point and pole singularities. If $z>0$ the contour of integration is deformed to the upper--half plane $\Im k_{z}>0$ (see Figure \ref{figura2}) whereas, if $z<0$ the contour of integration is deformed to the lower--half plane $\Im k_{z}<0$. The integration along the branch cut $B_1$ results in a   volume wave  \cite{wait,michalski}, which consist of a continuous spectrum of radiation modes.
 Here, we focus on distances between emitters of the same order than the propagation length of the graphene wire  eigenmodes, thus 
the intensity of the electric field reached by the excitation of these modes dominates the energy transfer process and the volume wave mode contribution can be neglected.  Then, the application of the residues theorem to Eqs. (\ref{camporho}), (\ref{campophi}) and (\ref{E2z}) gives the following components of the scattered fields
\begin{eqnarray}\label{camporho1}
E^{(2)}_{\rho}(\textbf{r})= 2 \pi i 
 \sum_{m=-\infty}^{+\infty} e^{i m \phi} e^{i \alpha_m z} \nonumber\\
\times \left[
-\frac{k_0 m}{\gamma_m^{(2)}} \frac{H_m(\gamma_m^{(2)}\rho)}{\gamma_m^{(2)}\rho} \mbox{Res} \, a_m^{(2)} + i \frac{\alpha_m}{\gamma_m^{(2)}} H'_m(\gamma_m^{(2)}\rho) \mbox{Res}\, b_m^{(2)}\right],    
\end{eqnarray}
\begin{eqnarray}\label{campophi1}
{E}^{(2)}_{\phi}(\textbf{r})=- 2\pi i
\sum_{m=-\infty}^{+\infty}\ e^{i m \phi} e^{i \alpha_m z}\nonumber \\
\times \left[
i\frac{k_0}{\gamma_m^{(2)}} H'_m(\gamma_m^{(2)}\rho) \mbox{Res}\, a_m^{(2)} +  \frac{\alpha_m m}{\gamma^{(2)}} \frac{H_m(\gamma_m^{(2)}\rho)}{\gamma_m^{(2)} \rho} \mbox{Res} \, b_m^{(2)}\right],
\end{eqnarray}
\begin{eqnarray}\label{campoz1}
E^{(2)}_z(\textbf{r})=2 \pi i \sum_{m=-\infty}^{+\infty}  e^{i m \phi} e^{i \alpha_m z}  H_m(\gamma_m^{(2)} \rho) \mbox{Res} \, b^{(2)}_m, 
\end{eqnarray}
where $\alpha_m$ is the propagation constant of a particular eigenmode, $\gamma_m^{(2)}=\sqrt{k_0^2 \varepsilon_2-\alpha_m^2}$ and Res is the residue of the integrand in (\ref{camporho}), (\ref{campophi}) and (\ref{E2z}) at the pole $\alpha_m$. 
\begin{eqnarray}\label{Res}
\mbox{Res}\, a_m^{(2)}= \lim_{k_z \to \alpha_m}(k_z-\alpha_m) a_m^{(2)}, \nonumber \\
\mbox{Res}\, b_m^{(2)}= \lim_{k_z \to \alpha_m}(k_z-\alpha_m) b_m^{(2)}. \nonumber \\
\end{eqnarray}
Inserting the expressions for $\textbf{E}^{(2)}(\textbf{r})=E_{\rho}^{(2)}\hat{\rho}+E_{\phi}^{(2)}\hat{\phi}+E_{z}^{(2)}\hat{z}$ into Eq. (\ref{ETn}) we obtain the following expression for the normalized energy transfer rate,
\begin{eqnarray}\label{ETnmodos}
F_{ET}=\sum_{mn} F_{nm} = \sum_n F_{nn}+\sum_{n<m} (F_{nm}+F_{mn}), 
\end{eqnarray}
where 
\begin{eqnarray}\label{ETnmodo}
F_{mn}=\frac{\hat{n}_A\cdot \textbf{E}_m(\textbf{r}_A) \, \hat{n}_A\cdot \textbf{E}_n^*(\textbf{r}_A) }{|\hat{n}_A\cdot \textbf{E}_0(\textbf{r}_A)|^2}.
\end{eqnarray}
%
The terms in the first sum in Eq. (\ref{ETnmodos}) represent the  contribution of each of the $n$ eigenmode channels to the normalized energy transfer rate  whereas each term in the second sum  
arises from the interference between $m$ and $n$ mode channels, 
\begin{equation}\label{Flj}
I_{nm}=F_{n,m}+F_{m,n} \approx \cos [(\Re \alpha_{\mbox{\tiny{n}}}-\Re \alpha_{\mbox{\tiny{m}}}) z].
\end{equation}         

\section{Results}

In this section, we apply the formalism sketched in previous section to calculate the energy transfer rate between two
emitters localized close to a graphene--coated wire. In all the calculations we assume that  the core is made of a transparent material with constitutive parameters $\varepsilon_1 = 2.13,\,\mu_1=1$ (corresponding to Polymethylpentene) and is embedded in vacuum ($\varepsilon_2 = \mu_2 = 1$). 
 In order to obtain separate contributions of different SP modes, firstly the propagation constant of these modes are obtained by requiring the denominator in the amplitudes $a_m^{(2)}$ and $b_m^{(2)}$ to be zero (see \ref{ap2}),
\begin{equation}\label{dispersion2}
\begin{array}{ll}
D_{m}=\left[-g_m+\frac{4 \pi}{c}\sigma \frac{m k_z}{(\gamma^{(2)})^2 a} j_m(x_1)\right]
 \left[g_m-\frac{4 \pi}{c} \sigma \frac{m k_z}{(\gamma^{(1)})^2 a} h_m(x_1)\right] \\
-\left[k_0(h_m(x_2)-j_m(x_1))+\frac{4\pi}{c}\sigma i k_0^2 a j_m(x_1) h_m(x_2)\right]\\ 
\times \left[k_0(\varepsilon_2 h_m(x_2)-\varepsilon_1 j_m(x_1))+\frac{4\pi}{c}\sigma i (\frac{1}{a}+\frac{m^2 k_z^2}{(\gamma^{(1)} \gamma^{(2)})^2 a^3})\right] = 0,
\end{array}
\end{equation}
where $x_j=\gamma^{(j)}a$ ($j=1,\,2$) and $j_m(x)=\frac{J'_m(x)}{x J_m(x)}$, $h_m(x)=\frac{H'_m(x)}{x H_m(x)}$. This condition is the dispersion relation of SPs and it determines the complex propagation constant $k_z=\alpha_m$ in terms of all the parameters of the graphene--coated wire. Equation (\ref{dispersion2}) has complex solutions with the same modulus and opposite signs, corresponding to SP propagation along the directions $\pm{z}$. We have found the complex roots of equation (\ref{dispersion2}) by adapting a numerical code  based on Newton Raphson method. In our examples we have selected the solution with $\Re\,\alpha_m > 0$ corresponding to wave propagation along the $+z$ direction.

In the  $m = 0$ case, Eq. (\ref{dispersion2})  is decoupled into two independent equations: an equation whose solutions
 have the longitudinal component of the magnetic field $H_z = 0$ (TM$z$ polarization),
\begin{eqnarray}\label{dispersionTMz}
\varepsilon_2 h_0(x_2)-\varepsilon_1 j_0(x_1)+\frac{4\pi}{c}\sigma i \frac{1}{k_0 a}= 0,
\end{eqnarray}
and another equation whose solutions have the longitudinal component of the electric field $E_z = 0$ (TE$z$ polarization),
\begin{eqnarray}\label{dispersionTEz}
h_0(x_2)-j_0(x_1)+\frac{4\pi}{c}\sigma i k_0 a j_0(x_1) h_0(x_2)= 0.
\end{eqnarray}

In the limit $a \rightarrow \infty$, taking into account the asymptotic expressions of the Bessel and Hankel functions \cite{abramowitz}, the dispersion equation (\ref{dispersionTMz}) converges to the TM dispersion equation of a single graphene sheet \cite{depineguias} 
\begin{eqnarray}\label{dispersionTM}
\frac{\gamma^{(1)}}{\varepsilon_1}+\frac{\gamma^{(2)}}{\varepsilon_2}+\frac{4 \pi \sigma}{c k_0}\frac{\gamma^{(1)}}{\varepsilon_1}\frac{\gamma^{(2)}}{\varepsilon_2}=0,
\end{eqnarray}
whereas Eq. (\ref{dispersionTEz}) converges to the TE dispersion equation of a single graphene sheet 
\begin{eqnarray}\label{dispersionTE}
\gamma^{(1)}+\gamma^{(2)}+\frac{4 \pi \sigma}{c k_0}\gamma^{(1)}\gamma^{(2)}=0.
\end{eqnarray}

In the general case $m\not=0$, the modes are not decomposed into independent polarizations, 
and to calculate the propagation constant of these modes we must solve Eq. ($\ref{dispersion2}$). 

\begin{figure}
\centering
\resizebox{0.45\textwidth}{!}
{\includegraphics{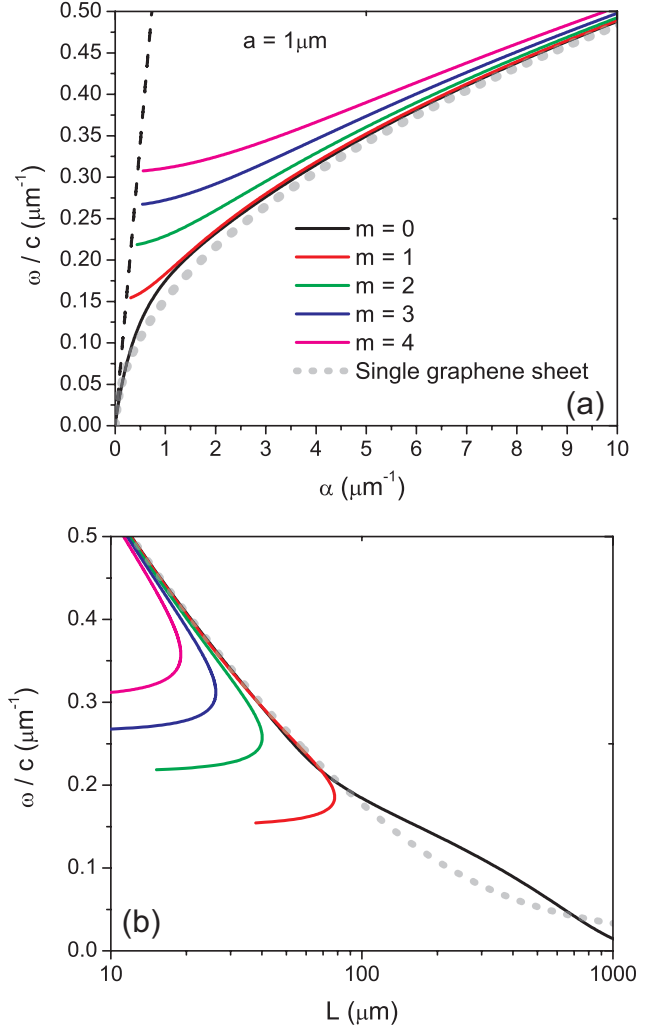}}
\caption{\label{fig:epsart} Dispersion curves for graphene--coated wire SP modes, calculated for $\mu_c=0.5$eV, $T=300$K, $\gamma_c=0.1$meV, $\varepsilon_1=2.13$ and $\varepsilon_2=1$. (a) $\Re\,\alpha_m$ and (b) $L=1/(2\Im\,\alpha_m)$ as a function of $\omega/c$. The black dashed line in (a) corresponds to the light line in medium 2. 
}\label{dispersion}
\end{figure}

As in the planar waveguides case, the energy transfer through the graphene--coated wire due to excitation of SPs is much greater than the corresponding energy transfer through the excitation of wave guided modes (modes which are evanescent outside the graphene--coated wire and standing waves in the insulator core) \cite{cuevas0,olivo}. This is true because graphene SP modes are highly localized in comparison with waveguide (or fiber) modes \cite{enguetaguias}. In addition, $s$--polarized SPs only exist for frequencies such as $\hbar \omega/ \mu_c>1.667$ where the imaginary part of the graphene conductivity  is  negative (see appendix \ref{grafeno}). For frequencies bellow this value, $\omega<1.667 \mu_c/\hbar$, (the frequency range studied in this work) only $p$--polarized SPs are well defined. Thus, as we have verified, the $p$--polarized SP modes contributions dominate the energy transfer process on the frequency region presented in figure \ref{dispersion} and consequently all other mode contributions,  which are solution of Eq. (\ref{dispersion2}), can be neglected in equation (\ref{ETnmodos}).

Figure \ref{dispersion} shows the propagation constant $\alpha_m$ ($0\leq m \leq 4$) of  SPs as a function of
the frequency $\omega/c$ for $a=1\mu$m.  The Kubo parameters are $\mu_c=0.5$eV, $\gamma_c=0.1$meV and $T=300$K. With this value of $\mu_c$, $p$--polarized SPs are supported by the graphene wire for frequency values less than $\omega/c \approx 4\mu$m$^{-1}$. These figures also show the curves corresponding to the propagation constant of SPs supported by a single graphene sheet interface (dotted line), \textit{i.e.}, a system with the flat graphene sandwiched between two dielectric half space with permittivities $\varepsilon_1$ and $\varepsilon_2$, calculated by solving Eq. (\ref{dispersionTM}).
From Figure \ref{dispersion}a we see that SP curves exist to the right of the light line of the inner dielectric medium. As in the metallic wire case \cite{dereux}, the $m=0$ mode goes down to $\omega/c=0$ approaching the light line from the right. The modes with $m>0$ intersects the light line at a finite $\omega/c$ value. Figure \ref{dispersion}b shows the propagation length $L$ of the SPs calculated as $L=\frac{1}{2\, \Im\,\alpha}$. For $\omega/c$ frequency values large enough,  
both functions $\Re \alpha_m(\omega/c)$ and $L(\omega/c)$  take a value which is almost equal to that of the single graphene sheet interface. 
\begin{figure}
\centering
\resizebox{0.45\textwidth}{!}
{\includegraphics{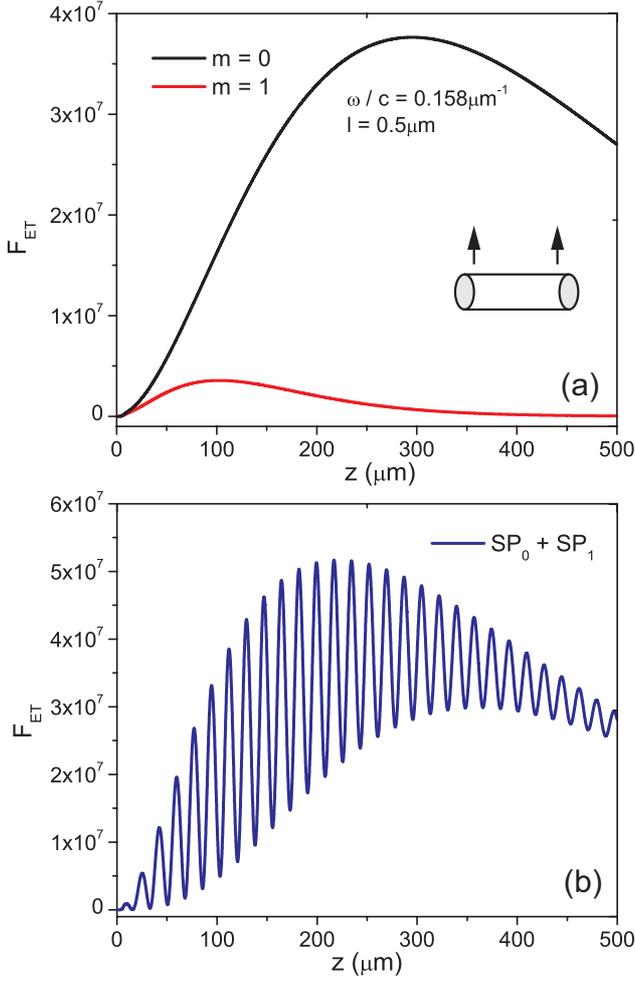}}
\caption{\label{fig:epsart} (a) Contribution of the SP$_0$ and the SP$_1$ ($F_{00}$ and $F_{11}$), to the normalized energy transfer rate between two dipoles, both placed at a distance $l=0.5\mu$m from the surface of the graphene wire ($\rho_D=\rho_A=1.5\mu$m, $\phi_D=\phi_A=0$, $z_D=0$), as a function of the separation $z$ along the wire. The frequency $\omega/c=0.185\mu$m$^{-1}$ and the dipole moments are oriented along the radial direction. (b) Normalized energy transfer rate as a superposition of the SP$_0$ and SP$_1$ modes as a function of the $z$ separation between the donor and acceptor. The waveguide parameters are the same as in figure \ref{dispersion}.
}\label{omegac0p158}
\end{figure}

Once the SP propagation constants are calculated, the contribution of each modes to the energy transfer rate between donor and acceptor can be obtained by using Eq. (\ref{ETnmodo}). To illustrate the interesting case where the coherent interference between these SP channels leads to large   spatial variations, we chose a frequency value for which two of the plasmonic bands plotted in Figure \ref{dispersion} show a significant difference in their propagation constant values.  
Figure \ref{omegac0p158} shows the contribution of SP$_0$ and SP$_1$ modes (SP modes with $m=0$ and $m=1$) to the energy transfer rate ($F_{mn}$ with $m=n=0$ and $m=n=1$) as a function of the separation $z$ between the donor and acceptor and for $\omega/c=0.158\mu$m$^{-1}$. The locations are $\textbf{r}_D=1.5\mu$m$\hat{\rho}$ and $\textbf{r}_A=1.5\mu$m$\hat{\rho}+z\hat{z}$ for the donor and acceptor, respectively, and both dipole moments are aligned along the $\rho$ direction (the directions of the dipole moments are shown in the graphs by the directions of the arrows). As can be seen in Figure \ref{dispersion}, only SPs with $m=0$ and $m=1$ are propagating and as a consequence only these modes carry energy through the graphene wire. Figure \ref{omegac0p158}a shows that the energy transfer contribution curves reach a maximum value for a certain value of the $z$ separation, $z_{max,m=0} \approx 300\mu$m for the SP$_0$ mode and $z_{max,m=1} \approx 100 \mu$m for the SP$_1$ mode. 
On the other hand, from Figure \ref{dispersion}b we determine the propagation lengths of SP modes resulting: $L_{m=0}=148.5\mu$m for the SP$_0$ mode and $L_{m=1}=51.5\mu$m for the SP$_1$ mode. This result suggests that the maximum contributions to the total normalized energy transfer rate are reached at a distance, between the donor and acceptor, two times larger than the SP propagation lengths.   
This fact can be understood  
as follows: due to the $\hat{\rho}$--component of the SP electric field (\ref{camporho1}) depends on the $z$ distance as $E_{z}(z) \approx e^{i \alpha_m z}=e^{i \Re \alpha_m z-\Im \alpha_m z}$, and the fact that in the absence the graphene wire, the field of the donor is written as (refer to \ref{ap1} for its derivation) $E_{0,z}(z) = \frac{e^{i k_2 z}}{z}$, it follows  that the energy transfer contribution of either of the two SP channels (\ref{camporho1}) can be written as
\begin{equation}
F_{SP} = |\frac{E_{\rho}(z) }{E_{0,z}(z) }|^2 =z^2 e^{-2 \Im \alpha_m z}, 
\end{equation}         
which reaches its maximum value at $z=\frac{1}{\Im \alpha_m}=2 L$. It is worth noting that, if 
the same donor--acceptor configuration (where both donor and acceptor dipole moments are oriented perpendicularly to the graphene surface) 
is 
placed near a planar graphene system, such as a single graphene sheet or a planar graphene waveguide, 
then the maximum energy transfer value is reached at a donor--acceptor separation which coincides with the SP propagation length \cite{olivo}. This difference lies in the geometry of the system through which the energy emitted by the donor dipole travels. As its required by energy conservation, in the planar graphene case (2D case), the plasmonic energy amplitude is proportional to $1/length$ ($length = \mbox{distance between the emission and the observation point}$) whereas for the wire case (1D case) this amplitude is constant.

Figure \ref{omegac0p158}b shows the energy transfer rate between the donor and acceptor as a coherent superposition of the SP$_0$ and the SP$_1$ contributions and calculated by using Eq. (\ref{ETnmodos}).   Due to the presence of the interference term $I_{nm}$ with $n=0$ and $m=1$,   this curve shows a pronounced spatial oscillation whose period $\Lambda=17.2\mu$m, a value that can be calculated within the framework of the model in Eq. (\ref{Flj}), where the interference term between the SP$_0$ and SP$_1$ is written as   
\begin{equation}\label{Ilj}
I_{0,1}=F_{0,1}+F_{1,0} \approx \cos [(\Re \alpha_{\mbox{\tiny{1}}}-\Re \alpha_{\mbox{\tiny{0}}}) z].
\end{equation}         
Equation (\ref{Flj}) shows a periodic spatial dependence along $z$ direction with a period  $\Lambda=2 \pi / (\Re \alpha_{\mbox{\tiny{0}}}-\Re \alpha_{\mbox{\tiny{1}}})=2 \pi/(0.794\mu\mbox{m}^{-1}-0.428\mu\mbox{m}^{-1}) \approx 17.2\mu$m, where we have used the values of the  SP propagation constants calculated in Figure \ref{dispersion}a for  $\omega/c=0.158\mu$m$^{-1}$. 
\begin{figure}[htbp!]
\centering
\resizebox{0.45\textwidth}{!}
{\includegraphics{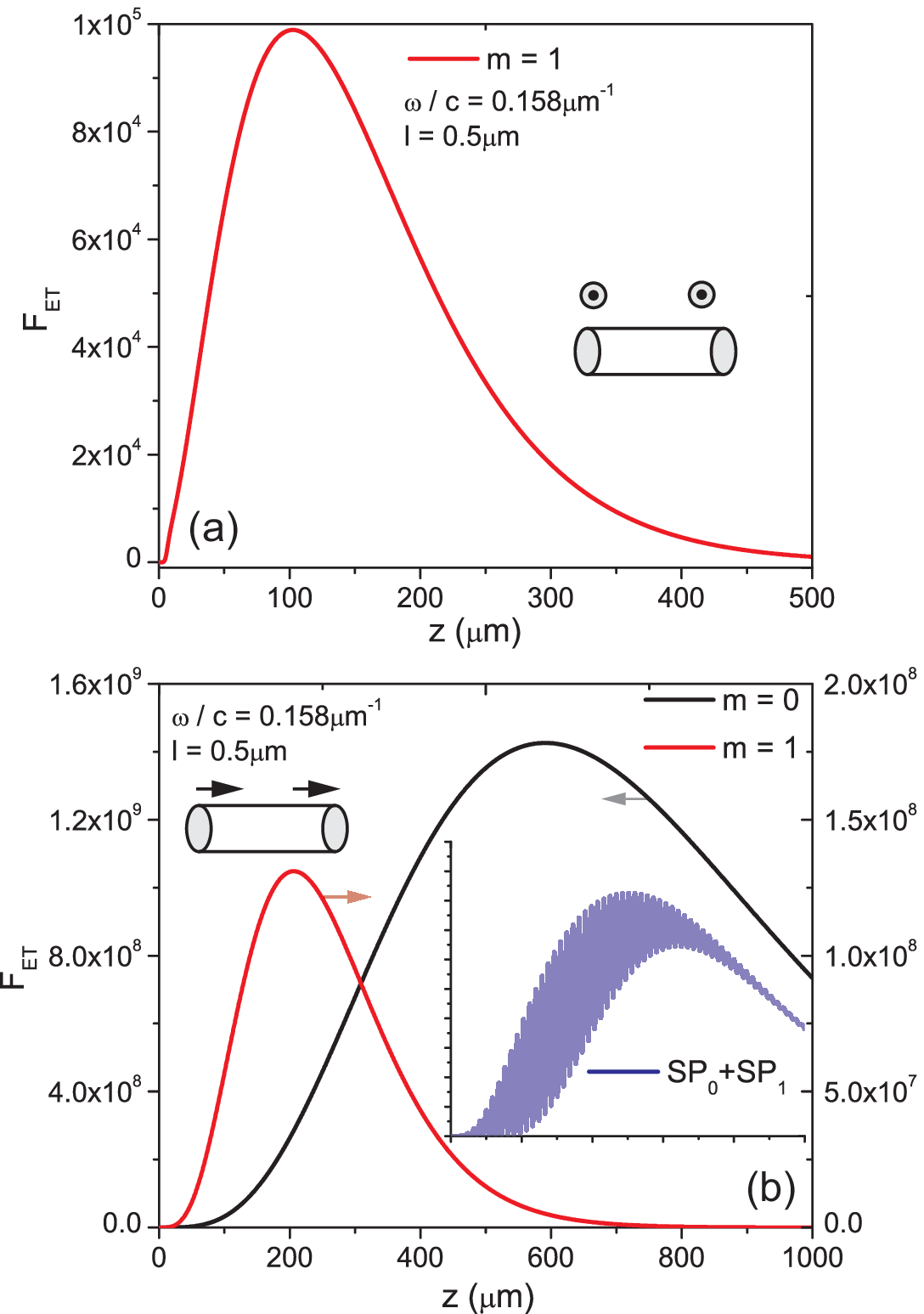}}
\caption{\label{fig:epsart} Contribution of the SP$_0$ and the SP$_1$ ($F_{00}$ and $F_{11}$), to the normalized energy transfer rate between two dipoles, both placed at a distance $l=0.5\mu$m from the surface of the graphene wire ($\rho_D=\rho_A=1.5\mu$m, $\phi_D=\phi_A=0,\, \mbox{and}\,\, z_D=0$), as a function of the separation $z$ along the wire. The frequency $\omega/c=0.185\mu$m$^{-1}$ and the dipole moments are oriented along the $\hat{z}$ direction (a) and along $\hat{\phi}$ direction (b). The inset in (b) shows the total  energy transfer rate as a superposition of the SP$_0$ and SP$_1$ modes as a function of the $z$ separation between the donor and acceptor. The waveguide parameters are the same as in figure \ref{dispersion}.
}\label{omegac0p158yy_zz}
\end{figure}

We next consider the case where the two dipole moments are parallel to the surface of the cylinder but  perpendicular to the axis of the cylinder, $\hat{p}_D=\hat{p}_A=\hat{\phi}$ as shown in the inset in Figure \ref{omegac0p158yy_zz}a.  
Since the electric field component of the SP$_0$  along the $\hat{\phi}$ direction is zero \cite{depineguias}, the $m=0$ plasmonic order is not coupled to the dipole incident field, and as a consequence the energy transfer contribution associated to this order is zero [we have numerically verified this assertion]. In Figure \ref{omegac0p158yy_zz}a we have plotted  the energy transfer contribution of the SP$_1$ mode. As in the previously presented
case where the donor and acceptor dipole moments are perpendicular to the graphene cylinder, we also observe that the maximum normalized energy transfer rate is obtained for a value of the $z$ separation two times larger than  the propagation length of the SP, $z_{max} \approx 100\mu$m$=2L_{m=1}$.  

In Figure \ref{omegac0p158yy_zz}b we have plotted the energy transfer contribution for the case where the two dipole moments are parallel to the axis of the cylinder, $\hat{p}_D=\hat{p}_A=\hat{z}$. Unlike previous configurations, the
distance $z_{max}\approx 600\mu$m for the $m=0$ order and $z_{max}\approx 200\mu$m for the $m=1$ order. These  values represent fourfold the SP propagation length of the $m=0$ and $m=1$ orders. 
This fact can be understood by taking into account that, in the absence of the graphene cylinder, the $z$  component of the donor electric field for $x_A=x_D$ is written as (see \ref{ap1}) 

\begin{equation}
E_{0,z}(z) \approx \frac{e^{i k_1 z}}{z^2}, 
\end{equation} 
while the $\hat{z}$ component of the SP electric field is as in Eq. (\ref{E2z}), $E_{z}(z) \approx e^{i \alpha_m z}=e^{i \Re \alpha_m z-\Im \alpha_m z}$. Thus,  the energy transfer contribution of either of the two SP channels (\ref{campoz1}) can be written as
\begin{equation}
F_{SP} = |\frac{E_{\rho}(z) }{E_{0,z}(z) }|^2 =z^4 e^{-2 \Im \alpha_m z}, 
\end{equation}         
which reaches its maximum value at $z=\frac{2}{\Im \alpha_m}=4 L_m$. 
Note that the values of the normalized energy transfer rate shown in Fig. \ref{omegac0p158yy_zz}b (donor--acceptor emitters with dipole moment oriented parallel to the wire axis) are  approximately two orders of magnitude larger than that corresponding to perpendicular orientation plotted in Figs. \ref{omegac0p158} and  \ref{omegac0p158yy_zz}a. This is true because the energy transfer rate without graphene wire decays as $z^{-4}$ for donor--acceptor emitters oriented parallel to the wire axis  whereas it decay as $z^{-2}$ for other orientations of the emitters.
 In the inset in Figure \ref{omegac0p158yy_zz}b we plotted the curve corresponding to the energy transfer rate as a superposition of the SP$_0$ and SP$_1$ modes as a function of the $z$ separation between the donor and acceptor. This curve presents an oscillation whose spatial  period is the same as in Figure \ref{omegac0p158}b, $\Lambda=17.2\mu$m.   

%
%

It is worth noting that the calculation of the energy transfer for small donor--acceptor separations ($z \approx \lambda=2 \pi c/\omega$) is a limitation of the presented model.  At these distances, the contribution of the radiative modes, which are excluded in our model, is not negligible and  our calculations depart significantly from experimental values.  Since  the emitter--SP coupling decreases with the emitter--cylinder distance, we expect this effect becomes even more pronounced when emitters are far away from the surface of the cylinder.

\section{Conclusions} \label{conclusiones}

In conclusion, we have examined the energy transfer rate behavior from a donor to an acceptor close to a graphene--coated  wire of circular cross section. We developed an analytical classical
method and obtained a rigorous solution for the electromagnetic field  in an integral form.  By solving the eigenmodes dispersion equation, which is related to the study of the non--trivial solutions to the boundary value problem in the absence of external sources, we have obtained  the complex propagation constants of the  graphene--wire SPs. 
This allowed  us to calculate the contribution of each of these SPs  to the total energy transfer rate. 

In all the presented examples, we have provide a comprehensive analysis about the strong impact of SPs on the energy transfer rate between two single emitters in terms of their kinematic characteristics. 
Our calculations show that the SP maximum contribution to the total normalized energy transfer is reached when the donor--acceptor distance is twice the value of the SP propagation length for the configuration when both dipole moments are oriented perpendicular to the wire axis. This result differ from that obtained for planar graphene systems, such as a single graphene sheet or a planar graphene waveguide for which the maximum energy transfer value is reached at a donor--acceptor separation which coincides with the SP propagation length.   
In case where the dipole moments of both the donor and acceptor are parallel to the wire axis, these maximum contributions are reached at a distance four times larger than the SP propagation lengths.

In addition, we have obtained a simple relation, which uses the SP dispersion relation, that allows us to calculate the spatial dependence of the normalized energy transfer rate between two quantum emitters. 
Although in our examples only it is consider three polarization directions, the  method described in section \ref{teoria} allows to calculate the energy transfer regarding an arbitrary dipole moment orientation in relation to the graphene--coated wire. In particular, from the presented  results it can be inferred that, for an arbitrary orientation of the dipole moments, each of the modal plasmonic contributions to the normalized energy transfer rate will have two characteristic  lengths related with the propagation length of the SPs involved. 

 Finally, the research can be continued by considering graphene--coated wire with a finite lenght. 
In this case,  reflections at the wire facets modify the dispersion relation of SPs playing a key role in the energy transfer process. A simple semi--analytical  approach used in case of a metallic wire  \cite{lalane}   considers the field confinement in the transverse wire direction and reflections at the wire ends as two completely independent problems. This procedure  leads to dispersion curves  of an infinite cylinder but with an intermode spacing which is proportional to the inverse wire length (Fabry--Perot modes).  A most rigorous approach that could be applied to the graphene--coated wire takes into account the formulation of the equivalence theorem to obtain a dispersion relation of the Fabry--Perot modes \cite{bordo}. As a result of applying these two approaches, the energy transfer rate would be enhanced at the frequencies corresponding to Fabry--Perot modes. Although we could planning such study in a future, as a first step, we believe that our contribution can be valuable for experimentalist, which using the dispersion relation of SPs, can calculate the interaction distance between a pair of quantum  emitters by  using a simple relation.

\section*{Acknowledgment} 
The authors acknowledge the financial support of Consejo Nacional de Investigaciones Cient\'{\i}ficas y T\'ecnicas (CONICET). Discussions with Vasilios Karanikolas (Materials Science Department, School of Natural Sciences, University of Patras, Patras, Greece) are gratefully acknowledged.


\appendix 

\section{Electric field of a dipole  in presence of a graphene--coated wire} 
\label{ap2}
\setcounter{equation}{0}
\renewcommand{\theequation}{B{\arabic{equation}}}

We consider an electric dipole $\textbf{p}$ localized at $\textbf{r}_D=\rho_D \hat{\rho}+ \phi_D\hat{\phi}+ z_D\hat{z}$, at a distance $\rho_D>a$ from the center of the cylindrical wire. The current density of the electric dipole is
\begin{eqnarray}\label{corriente}
\textbf{j}(\textbf{r})=-i \omega \textbf{p} \, \delta(\textbf{r}-\textbf{r}_D)\nonumber\\
=-i \omega \textbf{p} \, \frac{1}{\rho} \delta(\rho-\rho_D) \delta(\phi-\phi_D) \delta(z-z_D).
\end{eqnarray} 
In an unbounded medium (constitutve parameters $\varepsilon_2$, $\mu_2=1$), the dipole vector potential is 
%
\begin{eqnarray}\label{A}
\textbf{A}^{(2)}(\textbf{r})=-i k_0 \textbf{p} \frac{e^{i k_2 |\textbf{r}-\textbf{r}_D|}}{|\textbf{r}-\textbf{r}_D|}.
\end{eqnarray}
where $k^2=\varepsilon_2 k_0^2$, $k_0$ is the modulus of the photon wave vector in vacuum, $\omega$ is the angular frequency, $c$ is the vacuum speed of light, $\varepsilon_2$ is the electric permittivity 
and the superscript denotes medium 2. In order to take advantage of the translational symmetry of the system along $z$ direction, we expand Eq. (\ref{A}) into cylindrical  eigenfunctions, 
\begin{eqnarray}
A_\tau^{(2)}(\rho,\phi,z)=\sum_{m=-\infty}^{+\infty} \int_{-\infty}^{\infty} dk_z e^{i k_z (z-z_D)} e^{i m (\phi-\phi_D)} \nonumber \\
\times J_m(\gamma^{(2)} \rho_<) H_m(\gamma^{(2)} \rho_>) \frac{k_0 p_\tau}{2},
\end{eqnarray} 
where the subscript $\tau$ denotes $\rho$, $\phi$ or $z$, $\gamma^{(2)}=\sqrt{k_0^2\varepsilon_2-k_z^2}$,            $\rho_<$ ($\rho_>$) is the smaller (larger) of $\rho$ and $\rho_D$, $p_{\tau}$ is the $\tau$ component of the electric dipole  and $J_m$ and $H_m$ are the nth Bessel and Hankel functions of the first kind, respectively.  By using the equations 
\begin{eqnarray}\label{H}
\textbf{H}^{(2)}(\rho,\phi,z)=\nabla \times \textbf{A}^{(2)}(\rho,\phi,z),
\end{eqnarray} 
\begin{eqnarray}\label{E}
\textbf{E}^{(2)}(\rho,\phi,z)=\frac{i}{k_0 \varepsilon_2}\nabla \times \textbf{H}^{(2)}(\rho,\phi,z),
\end{eqnarray}
explicit expressions for the 
electric and magnetic fields emitted by the dipole are obtained 
%
\begin{eqnarray}\label{H2}
\textbf{H}^{(2)}_{inc}(\rho,\phi,z)=\sum_{m=-\infty}^{+\infty} \int_{-\infty}^{+\infty} dk_z e^{i k_z z} e^{i m \phi} \nonumber\\
\left[h^{\rho}_m(\rho) \hat{\rho}+h^{\phi}_m(\rho) \hat{\phi}+ h^z_m(\rho) \hat{z}\right],
\end{eqnarray}
\begin{eqnarray}\label{E2}
\textbf{E}^{(2)}_{inc}(\rho,\phi,z)=\sum_{m=-\infty}^{+\infty} \int_{-\infty}^{+\infty} dk_z e^{i k_z z} e^{i m \phi} \nonumber\\
\left[e^{\rho}_m(\rho) \hat{\rho}+e^{\phi}_m(\rho) \hat{\phi} + e^z_m(\rho) \hat{z}\right].
\end{eqnarray}
For $\rho<\rho_D$, the complex amplitudes of the tangential fields $e_{m\,j}$ and $h_{m\,j}$ ($j=z,\,\phi$) are given by
\begin{equation}\label{hz}
\begin{array}{ll}
h_{m\,z}(\rho)=\frac{k_0 \gamma^{(2)}}{4 i} J_m(\gamma^{(2)}\rho) \\
\times \left[p_{-} e^{i \phi_D} H_{m-1}(\gamma^{(2)}\rho_D)+p_{+} e^{-i \phi_D} H_{m+1}(\gamma^{(2)}\rho_D) \right] \\ 
\times e^{-i m \phi_D-i k_z z_D},
\end{array}
\end{equation}
\begin{equation}\label{hphi}
\begin{array}{ll}
h_{m\,\phi}(\rho)=\frac{i k_0}{4}  [ k_z p_{-} e^{i \phi_D} J_{m-1}(\gamma^{(2)}\rho) H_{m-1}(\gamma^{(2)}\rho_D)  \\
+ k_z p_{+} e^{-i \phi_D} J_{m+1}(\gamma^{(2)}\rho) H_{m+1}(\gamma^{(2)}\rho_D)-\\
2  p_z \gamma^{(2)} J_m(\gamma^{(2)}\rho) H'_m(\gamma^{(2)} \rho_D) ] 
 e^{-i m \phi_D-i k_z z_D},
\end{array}
\end{equation}
\begin{equation}\label{ez}
\begin{array}{ll}
e_{z\,m}(\rho)=\frac{\gamma^{(2)}}{4 \varepsilon_2} J_m(\gamma^{(2)}\rho)  
 [ k_z p_{-} e^{i \phi_D} H_{m-1}(\gamma^{(2)}\rho_D)\\
- k_z p_{+} e^{-i \phi_D} H_{m+1}(\gamma^{(2)}\rho_D)+2 i p_z (\gamma^{(2)})^2 H_m(\gamma^{(2)} \rho_D) ]  \\ 
\times e^{-i m \phi_D-i k_z z_D},
\end{array}
\end{equation}
\begin{equation}\label{ephi}
\begin{array}{ll}
e_{m\,\phi}(\rho)=  [ p_{-} e^{i \phi_D} J_{m-1}(\gamma^{(2)} \rho) \{\frac{\gamma^{(2)} m}{4 \varepsilon_2 \rho} H_{m}(\gamma^{(2)}\rho_D)-\\
\frac{k_0^2}{4}H_{m-1}(\gamma^{(2)}\rho_D)\}+
 p_{+} e^{-i \phi_D} J_{m+1}(\gamma^{(2)} \rho)\\
\times \{-\frac{\gamma^{(2)} m}{4 \varepsilon_2 \rho} H_{m}(\gamma^{(2)}\rho_D)+\frac{k_0^2}{4}H_{m+1}(\gamma^{(2)}\rho_D) \}  - \nonumber \\
 \frac{i p_z k_z m}{2 \varepsilon_2 \rho} J_{m}(\gamma^{(2)} \rho)  H_m(\gamma^{(2)} \rho_D) ]  e^{-i m \phi_D-i k_z z_D},
\end{array}
\end{equation}
where the prime denotes the derivative with respect to the argument and $p_{\pm}=p_x \pm i p_y$. For $\rho>\rho_D$ we need to exchange $J_m(x)$ and $H_m(x)$ in equations (\ref{hz}) to (\ref{ephi}). 

The $\hat{z}$ component of the scattered electric and magnetic  fields, denoted by $E^{(j)}_z$ and $H^{(j)}_z$ ($j=1,\,2$), are expanded as a functions of cylindrical harmonics, one for the internal region ($\rho < a$, superscript $1$) and another one for the external region ($\rho > a$, superscript $2$),
\begin{eqnarray}\label{H1z}
H^{(1)}_z(\rho,\phi,z)=\sum_{m=-\infty}^{+\infty} \int_{-\infty}^{+\infty} dk_z e^{i m \phi} e^{i k_z z} a^{(1)}_m J_m(\gamma^{(1)} \rho), 
\end{eqnarray}
\begin{eqnarray}\label{E1z}
E^{(1)}_z(\rho,\phi,z)=\sum_{m=-\infty}^{+\infty} \int_{-\infty}^{+\infty} dk_z e^{i m \phi} e^{i k_z z} b^{(1)}_m J_m(\gamma^{(1)} \rho), 
\end{eqnarray}
\begin{eqnarray}\label{H2z}
H^{(2)}_z(\rho,\phi,z)=\sum_{m=-\infty}^{+\infty} \int_{-\infty}^{+\infty} dk_z e^{i m \phi} e^{i k_z z} a^{(2)}_m H_m(\gamma^{(2)} \rho), 
\end{eqnarray}
\begin{eqnarray}\label{E2z}
E^{(2)}_z(\rho,\phi,z)=\sum_{m=-\infty}^{+\infty} \int_{-\infty}^{+\infty} dk_z e^{i m \phi} e^{i k_z z} b^{(2)}_m H_m(\gamma^{(2)} \rho), 
\end{eqnarray}
where $a^{(j)}_m$ and $b^{(j)}_m$ ($j=1,\,2$) are unknown complex coefficients. Because of the cylindrical geometry, by combining the Faraday and  Ampere--Maxwell equations \cite{jackson} the transverse components of the electromagnetic fields ($\textbf{E}_{t}=E_{\rho}\hat{\rho}+E_{\phi} \hat{\phi},\,\textbf{H}_{t}=H_{\rho} \hat{\rho}+H_{\phi}\hat{\phi}$) can be obtained from their longitudinal components $E_{z}$ and $H_{z}$,
\begin{eqnarray}\label{campos_transversales}
\textbf{E}^{(j)}_t(\rho,\phi,z)=\frac{1}{(\gamma^{(j)})^2} \left[i k_z \nabla_t E^{(j)}_z-i k_0 \hat{z}\times \nabla_t H^{(j)}_z \right] \nonumber\\
\textbf{H}^{(j)}_t(\rho,\phi,z)=\frac{1}{(\gamma^{(j)})^2} \left[i k_z \nabla_t H^{(j)}_z+i k_0 \varepsilon_j \hat{z}\times \nabla_t E^{(j)}_z \right], 
\end{eqnarray}
where $\nabla_t=\hat{\rho}\frac{\partial}{\partial \rho}+\hat{\phi} \frac{1}{\rho} \frac{\partial}{\partial \phi}$ is the transverse part of the $\nabla$ operator. 
Due to the graphene coating, the tangential components of the magnetic field are no longer continuous across the boundary as they were in the case of uncoated cylinders. Considering this, the boundary conditions on the interface $\rho=a$ impose that
\begin{equation}\label{cc0}
\begin{array}{ll}
E_z^{(1)}|_{\rho=a}=E_z^{(2)}|_{\rho=a}+E_{inc\,z}|_{\rho=a} \\
&\\
E_{\phi}^{(1)}|_{\rho=a}=E_{\phi}^{(2)}|_{\rho=a}+E_{inc\,\phi}|_{\rho=a} \\
&\\
H^{(2)}_z|_{\rho=a}+H_{inc\,_z}|_{\rho=a}-H^{(1)}_z|_{\rho=a}\\
=-\frac{4 \pi}{c} \sigma E^{(2)}_{\phi}|_{\rho=a}\\
&\\
H^{(2)}_{\phi}|_{\rho=a} + H_{inc\,\phi}|_{\rho=a}-H_{\phi}|_{\rho=a}\\
&\\
=\frac{4 \pi}{c} \sigma E^{(2)}_{z}|_{\rho=a}.
\end{array}
\end{equation} 
Inserting the expressions of the tangential components of the incident and scattered fields into the boundary conditions, we obtain a system of equations for the amplitudes of the scattered fields,
\begin{equation}\label{cc}
\begin{array}{ll}
b^{(1)}_m J_m(x_1)=b^{(2)}_m h_m(x2)+e_{m\,z}(a) \\
&\\
\frac{m k_z}{(\gamma^{(1)})^2 a} b_m^{(1)} J_m(x_1)+\frac{i k_0}{(\gamma^{(1)})^2} a_m^{(1)} J'_m(x_1)= 
\frac{m k_z}{(\gamma^{(2)})^2 a} b_m^{(2)} H_m(x_2)+\\
\frac{i k_0}{(\gamma^{(2)})^2} a_m^{(2)} H'_m(x_2)-e_{m\,\phi}(a) \\
&\\
h_{m\,z}(a)+a_m^{(2)} H_m(x_2)-a_m(x_1) J_m(x_1)=\\
-\frac{4 \pi}{c} \sigma [-\frac{m k_z}{(\gamma^{(2)})^2 a} b_m^{(2)} H_m(x_2)-\frac{i k_0 }{\gamma^{(2)}} a_m^{(2)} H'_m(x_2) ] \\
&\\
h_{m\,\phi}(a)-\frac{m k_z}{(\gamma^{(2)})^2 a} a_m^{(2)} H_m(x_2)+\\
\frac{i k_0 \varepsilon_2}{\gamma^{(2)}} b_m^{(2)} H'_m(x_2)+\frac{m k_z}{(\gamma^{(1)})^2 a} a_m^{(1)} J_m(x_1)-\frac{i k_0 \varepsilon_1}{\gamma^{(1)}} b_m^{(1)} J'_m(x_1)=  \\
\frac{4 \pi}{c} \sigma b_m^{(2)} H_m(x_2)
\end{array}
\end{equation} 
where $x_1=\gamma^{(1)} a$ and $x_2=\gamma^{(2)} a$. It is  worth noting that the dipole field amplitudes $h_m^z$, $h_m^{\phi}$, $e_m^{z}$ and $e_m^{\phi}$ have been evaluated at $\rho=a$. Note that, in the case of $\sigma=0$, i.e., in the absence of current density induced in the graphene wrapping,  Eq. (\ref{cc}) converges to the standard boundary conditions  without graphene \cite{nha}.
The coefficients $a_m^{(1)}$ and $b_m^{(1)}$ are eliminated from Eqs. (\ref{cc}) and a system of equations for the coefficients $a_m^{(2)}$ and $b_m^{(2)}$ is obtained, 
\begin{equation}\label{matriz}
\left[
\begin{array}{ll}
M_{11} & M_{12}\\
M_{21} & M_{22}
\end{array} \right] 
\left( 
\begin{array}{ll}
a_m^{(2)}\\
b_m^{(2)}
\end{array} \right)=
\left( 
\begin{array}{ll}
c_m\\
d_m
\end{array} \right),
\end{equation}

%
%
%
where
\begin{equation}\label{cc2}
\begin{array}{ll}
M_{11}=k_0(h_m(x_2)-j_m(x_1))+\frac{4\pi}{c}\sigma i k_0^2 a j_m(x_1) h_m(x_2) \\
M_{12}=-g_m+\frac{4 \pi}{c}\sigma \frac{m k_z}{(\gamma^{(2)})^2 a} j_m(x_1) \\
M_{21}=-g_m+\frac{4 \pi}{c} \sigma \frac{m k_z}{(\gamma^{(1)})^2 a} h_m(x_1) \\
M_{22}=-\left[k_0(\varepsilon_2 h_m(x_2)-\varepsilon_1 j_m(x_1))+\frac{4\pi}{c}\sigma i (\frac{1}{a}+\frac{m^2 k_z^2}{(\gamma^{(1)} \gamma^{(2)})^2 a^3})\right],
\end{array}
\end{equation} 
\begin{equation}\label{cc2b}
\begin{array}{ll}
g_m=\frac{i m k_z}{a^2}\left[\frac{1}{(\gamma^{(2)})^2}-\frac{1}{(\gamma^{(1)})^2}\right]\\  
c_m=\frac{1}{H_m(x_2) i a} \left[\frac{m k_z}{(\gamma^{(1)})^2 a} e_m^{z}(a)+ e_m^{\phi}(a)\right]\\
+k_0 \frac{j_m(x_1)}{H_m(x_2)}\left[h_m^{z}(a)+\frac{4 \pi}{c}\sigma e_m^{\phi}(a)\right]\\
d_m=-\frac{1}{H_m(x_2) i a} \left[\frac{m k_z}{(\gamma^{(1)})^2 a} h_m^{z}(a)+h_m^{\phi}(a)\right]+ k_0\varepsilon_1 \frac{j_m(x_1)}{H_m(x_2)}e_m^{z}(a),
\end{array}
\end{equation} 
and
\begin{eqnarray}\label{jyh}
j_m(x)=\frac{J'_m(x)}{x J_m(x)}, \nonumber\\
h_m(x)=\frac{H'_m(x)}{x H_m(x)}.
\end{eqnarray}
By solving  Eq. (\ref{cc2}), explicit expressions for the complex amplitudes $a_m^{(2)}$ and $b_m^{(2)}$ are obtained
\begin{equation}\label{am}
a_m^{(2)}=\frac{c_m M_{22}-d_m M_{12}}{D_m},
\end{equation}
\begin{equation}\label{bm}
b_m^{(2)}=\frac{d_m M_{11}-c_m M_{21}}{D_m},
\end{equation} 
where $D_m$ is the determinant of the $2\times 2$ matrix in Eq. (\ref{matriz}).  

\section{Electric field of a dipole in an unbounding medium} 
\label{ap1}
\setcounter{equation}{0}
\renewcommand{\theequation}{A{\arabic{equation}}}

We consider an electric dipole at position $\textbf{x}_D$ oriented along $\hat{z}$ direction, $\textbf{p}=p\hat{z}$. The vector potential $\textbf{A}_e(\textbf{x})$ in spherical coordinates is given by \cite{novotny}
\begin{equation}\label{A_e_esfericas}
\textbf{A}_e(\textbf{x})=-i k_0\, \frac{ e^{i k_1 R}}{ R} p \hat{z},
\end{equation}  
where $R=|\textbf{x}-\textbf{x}_D|$. By using Eqs. (\ref{H}) and (\ref{E}), we obtain the following components for the electric field in cartesian coordinates

\begin{eqnarray}
\begin{array}{cc}
\label{Elibre}
E_{x}(\textbf{x})=
\frac{ e^{i k_1 R}}{\varepsilon_1 R} \left[\left(i k_1-\frac{1}{R}\right)^2+\left(-\frac{i k_1}{R}+\frac{2}{R^2}\right)\right]\frac{z-z_D}{R} \frac{x-x_D}{R},\\
E_{y}(\textbf{x})=
\frac{ e^{i k_1 R}}{\varepsilon_1 R} \left[\left(i k_1-\frac{1}{R}\right)^2+\left(-\frac{i k_1}{R}+\frac{2}{R^2}\right)\right]\frac{z-z_D}{R} \frac{y-y_D}{R},\\
E_{z}(\textbf{x})=
\frac{ e^{i k_1 R}}{\varepsilon_1 R} \\
\times \left\{\left[\left(i k_1-\frac{1}{R}\right)^2+\left(-\frac{i k_1}{R}+\frac{2}{R^2}\right)\right]\frac{(z-z_D)^2}{R^2}+\left(i k_1-\frac{1}{R}\right)\frac{1}{R}\right\}\\
+k_0^2 \frac{ e^{i k_1 R}}{R}.
\end{array}
\end{eqnarray}  
For the electric dipole orientation along the  $\hat{x}$ direction, $\textbf{p}=p\hat{x}$, the field components can be obtained from Eqs. (\ref{Elibre}) by replacing $E_x \rightarrow E_z$, $E_z \rightarrow E_x$, $x \rightarrow -z$ and $z \rightarrow x$. 

\section{Graphene conductivity} \label{grafeno}
\setcounter{equation}{0}
\renewcommand{\theequation}{C{\arabic{equation}}}
We consider the  graphene layer as an infinitesimally thin, local and isotropic two--sided layer with frequency--dependent surface conductivity $\sigma(\omega)$ given by the Kubo formula \cite{falko,milkhailov}, which can be read as  $\sigma= \sigma^{intra}+\sigma^{inter}$, with the intraband and interband contributions being
\begin{equation} \label{intra}
\sigma^{intra}(\omega)= \frac{2i e^2 k_B T}{\pi \hbar (\omega+i\gamma_c)} \mbox{ln}\left[2 \mbox{cosh}(\mu_c/2 k_B T)\right],
\end{equation}  
\begin{eqnarray} \label{inter}
\sigma^{inter}(\omega)= \frac{e^2}{\hbar} \Bigg\{   \frac{1}{2}+\frac{1}{\pi}\mbox{arctan}\left[(\omega-2\mu_c)/2k_BT\right]-\nonumber\\
\frac{i}{2\pi}\mbox{ln}\left[\frac{(\omega+2\mu_c)^2}{(\omega-2\mu_c)^2+(2k_BT)^2}\right] \Bigg\},
\end{eqnarray}  
where $\mu_c$ is the chemical potential (controlled with the help of a gate voltage), $\gamma_c$ the carriers scattering rate, $e$ the electron charge, $k_B$ the Boltzmann constant and $\hbar$ the reduced Planck constant.


\end{document}